\documentclass[aps,twocolumn,amsmath,amsfonts,nofootinbib,preprintnumbers,
  superscriptaddress,secnumarabic]{revtex4-1}
\usepackage[utf8]{inputenc}
\usepackage{color,graphicx,hyperref,amsmath,amssymb,multirow,slashed}
\hypersetup{colorlinks=true,linkcolor=blue,citecolor=magenta,filecolor=magenta,
  urlcolor=cyan}
\newcommand{\be}{\begin{equation}}
\newcommand{\ee}{\end{equation}}
\def\bsp#1\esp{\begin{split}#1\end{split}}
\def\bpm{\begin{pmatrix}}
\def\epm{\end{pmatrix}}
\def \met{\slashed{E}_T}
\def \ptmiss{\slashed{\bf p}_T}
\def \dch{H_1^{\pm\pm}}

\def \mdch{m_{H_1^{\pm\pm}}}

\def \mdchg{m_{\tilde{\chi}_1^{\pm\pm}}}
\def\ie{\textit{i.e.}}

\begin{document}
\preprint{{\scriptsize} HRI-RECAPP-2020-003, 
CUMQ/HEP 191, HIP-2020-7/TH}
\title{The left-right supersymmetric option at a high-energy upgrade of the LHC}

\author{Mariana Frank}
\email{mariana.frank@concordia.ca}
\affiliation{Department of Physics, Concordia University, 7141 Sherbrooke St.West, Montreal, QC, Canada H4B 1R6}
\author{Benjamin Fuks}
\email{fuks@lpthe.jussieu.fr}
\affiliation{Sorbonne Universit\'e, CNRS, Laboratoire de Physique Th\'eorique et Hautes \'Energies, LPTHE, F-75005 Paris, France, 
\& Institut Universitaire de France, 103 boulevard Saint-Michel, 75005 Paris, France}
\author{Katri Huitu}
\email{katri.huitu@helsinki.fi}
\affiliation{Department of Physics, and Helsinki Institute of Physics, P. O. Box 64, FI-00014 University of Helsinki, Finland}
\author{Subhadeep Mondal}
\email{subhadeep.mondal@helsinki.fi}
\affiliation{Department of Physics, and Helsinki Institute of Physics, P. O. Box 64, FI-00014 University of Helsinki, Finland}
\author{Santosh Kumar Rai}
\email{skrai@hri.res.in}
\affiliation{Regional Centre for Accelerator-based Particle Physics,
  Harish-Chandra Research Institute, HBNI, Chhatnag Road, Jhusi, Prayagraj 211019, India}
\author{Harri Waltari}
\email{h.waltari@soton.ac.uk}
\affiliation{Department of Physics and Astronomy, University of Southampton, Highfield, Southampton SO17 1BJ, United Kingdom, 
\& Department of Particle Physics, Rutherford Appleton Laboratory, Didcot OX11 0QX, United Kingdom}

\date{\today}

\begin{abstract}
We investigate the possibility that a minimal realization of left-right
supersymmetry can be reachable at a high-energy upgrade of the LHC, expected to
operate at a center-of-mass energy of 27~TeV. This minimal scenario
has a relatively light $SU(2)_R$ doubly-charged Higgs boson, which could decay dominantly
into tau-lepton pairs. We explore the associated signals comprised of at least three
hadronically-decaying taus, or with at least two hadronic taus and one
same-sign-same-flavor charged lepton pair. Our analysis shows that the former
signature is challenging to use for getting handles on the signal due to the large
corresponding background, and that the latter one can lead to a handful of new
physics events in an almost background-free environment. 
We however find that a signal comprised of three hadronically-decaying tau
leptons is likely to be observed at a low luminosity of proton-proton collisions
at a 27~TeV upgrade of the LHC.
\end{abstract}

\maketitle
\section{Introduction}
\label{sec:intro}

During the last years of operation at the Large Hadron Collider (LHC), no
significant deviation from the Standard Model (SM) predictions has been found.
Still, the SM as it stands is incomplete as it fails to explain, for instance,
neutrino masses, dark matter and the baryon-antibaryon asymmetry of the
Universe. However, new particles and interactions have failed (so far) to
materialize at the LHC. Of all the candidates for physics beyond the SM, weak
scale supersymmetry~\cite{Nilles:1983ge,Haber:1984rc} is amongst the most
promising ones. It associates one partner of opposite statistics with each of
the SM degrees of freedom and unifies the Poincar\'e symmetry with the internal
gauge symmetries. This setup leads to an elegant solution to some of
the limitations of the SM, as for instance concrete and realistic
supersymmetric realizations usually include a natural dark matter candidate as
the lightest supersymmetric particle (LSP).

On the other hand, left-right symmetric models~\cite{Pati:1974yy,
Mohapatra:1974hk,Mohapatra:1974gc} provide a natural mechanism to generate
neutrino masses. In this case, parity is a symmetry of the theory (that has
to be further broken dynamically) and the strong $CP$ problem is solved, thanks
to the enlarged gauge symmetry group~\cite{Mohapatra:1995xd,Kuchimanchi:1995rp,
Mohapatra:1996vg}.
In the supersymmetric context, in which both supersymmetry and left-right
motivations are combined, the same extended gauge symmetry reason leads
additionally to the automatic absence of any $R$-parity-violating interaction.
This therefore prevents the proton from being unstable and guarantees a viable
dark matter candidate as the LSP. However, the simplest left-right
supersymmetric realizations often predict upper bounds for particle
masses that do not easily agree with the latest non-observations of any hint for
new physics in LHC data, at least
when tree-level calculations are in order. Already the first proposal for a
left-right supersymmetric (LRSUSY) model hence suggested an $SU(2)_R$ charged
gauge boson with a mass satisfying $m_{W_R}\lesssim 1$~TeV~\cite{Cvetic:1983su},
which is today largely ruled out by the results of the LHC
experiments~\cite{Sirunyan:2016iap,Aaboud:2017yvp}. This has consequently led to
the development of LRSUSY models featuring an extended Higgs sector, so that the
$SU(2)_R$ boson masses could be pushed to a higher scale~\cite{Cvetic:1985zp}.

In this case, left-right symmetry breaking is often minimally built through
$SU(2)_R$ scalar triplets featuring two neutral, one singly-charged and one
doubly-charged Higgs degrees of freedom. At tree-level, the latter generally
acquires a vacuum expectation value (VEV) at the global minimum of the potential
that that corresponds to a configuration that is lower in energy than the
one in which only the neutral states develop a VEV. This problem can be cured
by invoking large
contributions either from non-renormalizable operators~\cite{Mohapatra:1995xd,
Mohapatra:1996vg,Aulakh:1998nn} or through loop corrections \cite{Babu:2008ep,
Frank:2011jia,Basso:2015pka}, or by spontaneous $R$-parity 
breaking~\cite{Kuchimanchi:1993jg}. The most appealing option consists of the
second one, in which loop corrections stabilize the charge-conserving
vacuum~\cite{Basso:2015pka}. In this setup, the model turns out to be
quite predictable, at least for what concern the properties of the $W_R$ boson
(and in particular its mass).

On different grounds, imposing a dark matter candidate with the right features,
that for instance leads to a relic density in agreement with Planck data,
further restricts the possibilities for the particle spectrum as the LSP has to
lie within some mass range below 1~TeV~\cite{Frank:2017tsm,Chatterjee:2018gca}.
With this phenomenologically constrained version of LRSUSY at hand, we
investigate in this work whether the future high-luminosity phase of the LHC
(HL-LHC) or its proposed 27~TeV energy upgrade, the so-called high-energy LHC
(HE-LHC)~\cite{Abada:2019ono}, could observe or rule out the model once and for
all.

The rest of this work is organized as follows. In Sec.~\ref{sec:model}, we
briefly describe our theoretical framework, detailing in particular how a
LRSUSY discovery at the HL-LHC could not happen. In Sec.~\ref{sec:collider}, we
focus on this most pessimistic case and design a set of representative
benchmark scenarios. We then demonstrate how the HE-LHC could provide
handles on the model. We summarize and conclude in Sec.~\ref{sec:conclusion}.

\section{Left-right supersymmetry}
\label{sec:model}

Left-right symmetric models~\cite{Pati:1974yy,Mohapatra:1974hk,Mohapatra:1974gc}
are based on the gauge group $SU(3)_C \times SU(2)_L \times SU(2)_R \times
U(1)_{B-L}$. The matter sector is defined by three families of left- and
right-handed quark and lepton supermultiplets,
\be\bsp
  & \hspace*{-.28cm}
   Q_L\!=\!\bpm u_L  \\ d_L\epm \!\!\sim\!
     ({\bf 3},      {\bf 2}, {\bf 1})_{\frac13}, \
   Q_R\!=\!\bpm d_R^c  \\-u_R^c\epm \!\!\sim\!
     ({\bf \bar 3}, {\bf 1}, {\bf 2}^*)_{-\frac13},\\
  & \hspace*{-.28cm}
   L_L\!=\!\bpm \nu_L\\ e_L\epm \!\!\sim\!
     ({\bf 1}, {\bf 2}, {\bf 1})_{-1}, \
   L_R\!=\!\bpm e_R^c  \\ -N^c \epm \!\!\sim\!
     ({\bf 1}, {\bf 1}, {\bf 2}^*)_1,
\esp\ee
where we include in our notation the representations of the various fields under
the LRSUSY gauge group, with the $U(1)_{B-L}$ charge given as a subscript.
Compared with the usual Minimal Supersymmetric Standard
Model (MSSM), the spectrum features a right-handed neutrino field $N$ (as part of
the $SU(2)_R$ lepton doublet $L_R$), so that Dirac neutrino mass term are
allowed in the superpotential. The Higgs sector of the model is quite rich and
includes an $SU(2)_R$ Higgs triplet  to break the $SU(2)_R \times
U(1)_{B-L}$ symmetry, its $SU(2)_L$ counterpart to preserve left-right parity,
as well as a pair of $SU(2)_L\times SU(2)_R$ Higgs bidoublets that are required
to generate all SM fermion masses. Neutrino masses
are hence generated through a combination of the Type-II~\cite{Magg:1980ut,
Cheng:1980qt,Mohapatra:1980yp,Lazarides:1980nt,Schechter:1980gr,Cai:2017mow} and
Type~I~\cite{Minkowski:1977sc,Yanagida:1979as,GellMann:1980vs,Glashow:1979nm,
Mohapatra:1979ia,Shrock:1980ct,Schechter:1980gr,Cai:2017mow} seesaw mechanisms, after the
breaking of the left-right symmetry. The Higgs sector moreover includes an
extra gauge singlet
that allows for the shift of the left-right symmetry breaking scale well beyond
the TeV regime. The Higgs superfield content and the associated representations
under the LRSUSY gauge group are hence summarized as
\be\bsp
\label{eq:higgs-decomp}
   \Phi_a = \bpm \Phi^+_{a1}&\Phi^0_{a1}\\ \Phi_{a2}^0& \Phi_{a2}^-\epm
   & \sim ({\bf 1},{\bf 2}, {\bf 2}^*)_0, \\
   \Delta_{L}=\bpm
      \frac {1}{\sqrt{2}}\Delta_L^-&\Delta_L^0\\
      \Delta_{L}^{--}&-\frac{1}{\sqrt{2}}\Delta_L^-
    \epm &\sim ({\bf 1}, {\bf 3}, {\bf 1})_{-2},\\
   \delta_{L}  = \bpm
    \frac {1}{\sqrt{2}}\delta_L^+&\delta_L^{++}\\
    \delta_{L}^{0}&-\frac{1}{\sqrt{2}}\delta_L^+
    \epm &\sim ({\bf 1}, {\bf 3}, {\bf 1})_2,\\
   \Delta_{R} = \bpm
    \frac {1}{\sqrt{2}}\Delta_R^-&\Delta_R^0\\
    \Delta_{R}^{--}&-\frac{1}{\sqrt{2}}\Delta_R^-
    \epm &\sim ({\bf 1}, {\bf 1}, {\bf 3})_{-2},\\
   \delta_{R}  = \bpm
    \frac {1}{\sqrt{2}}\delta_R^+&\delta_R^{++}\\
    \delta_{R}^{0}&-\frac{1}{\sqrt{2}}\delta_R^+
    \epm &\sim ({\bf 1}, {\bf 1}, {\bf 3})_2,  \\
    S\hspace*{2cm} &\sim ({\bf 1}, {\bf 1}, {\bf 1})_0 \ .
\esp\ee

The model Lagrangian includes, on top of usual gauge-invariant kinetic terms for
all fields, supersymmetric interaction terms originating from the
superpotential $W$ and a soft supersymmetry-breaking Lagrangian. The
superpotential reads
\be\label{superpotential}\bsp
  W = &\ Q_L^T {\bf Y}_{Q}^{(i)} \Phi_{i} Q_R +
        L_L^T {\bf Y}_{L}^{(i)} \Phi_{i} L_R +
        L_L^T {\bf h}_{LL} \delta_L L_L
  \\ &\ + L_R^T {\bf h}_{RR} \Delta_R L_R +
        \lambda_L\ S\ \mbox{Tr}\left[\Delta_L \delta_L\right]
  \\ &\ + \lambda_R\ S\ \mbox{Tr}\left[\Delta_R \delta_R\right] +
        \lambda_3\ S\ \mbox{Tr}\left[\tau_2 \Phi^T_1 \tau_2 \Phi_2\right]
  \\ &\ + \lambda_4\ S\ \mbox{Tr}\left[\tau_2 \Phi^T_1 \tau_2 \Phi_1\right]
    +  \lambda_5\ S\ \mbox{Tr}\left[\tau_2 \Phi^T_2 \tau_2 \Phi_2\right]
  \\&\ +\lambda_S\ S^3 + \xi_F\ S \ ,
\esp\ee
where the Yukawa couplings ${\bf Y}_{Q,L}$ and ${\bf h}_{LL,RR}$ are $3\times 3$
matrices in the flavor space, and the $\lambda$ and $\xi_F$ parameters are
associated with the various Higgs(ino) self-interactions. Moreover, in our
notation, $\tau_2$ denotes the second Pauli matrix and  we omit all indices for
clarity. We derive from the form of the superpotential the
corresponding soft terms, to which one should add scalar and gaugino mass terms.
Further details of the model can be found in refs.~\cite{Alloul:2013fra,
Frank:2017tsm}.

The superpotential possesses a $U(3,\mathbb{C})$ symmetry, 
whose spontaneous breaking leads to the appearance of several Goldstone bosons,
one of them being the $SU(2)_R$ doubly-charged Higgs
state~\cite{Babu:2008ep}. This doubly-charged state remains massless even after
the addition of soft SUSY breaking terms, and $D$-terms shifts its mass so that
it is nearly always tachyonic. Loop corrections however restores a positive
squared mass~\cite{Babu:2008ep,Basso:2015pka}. As the bulk of the mass is
loop-induced, the $SU(2)_R$ doubly-charged Higgs boson lies always in the
lightest part of the particle spectrum, in contrast with any other of the
numerous Higgs states of the model. Searches for the $SU(2)_R$
doubly-charged Higgs boson are therefore promising in either discovering
LRSUSY-related new physics, or excluding a large part of the parameter space.

The LRSUSY spectrum is largely determined by the scale of the left-right
symmetry breaking. While current experimental limits imply that this scale has
to lie in the multi-TeV range, it must at the same time satisfy an upper limit,
so that $\langle\Delta_R^0\rangle\sim v_R\in[10, 15]$~TeV
when all other parameters are held fixed~\cite{Basso:2015pka}.
Larger $v_R$ values would indeed destabilize the charge-conserving vacuum
configuration, as the scalar potential terms
\be\label{scalarpot}\bsp
  & \hspace{-.2cm}\Big|  \xi_F
   \!+\! \lambda_L \mbox{Tr}\left[\Delta_L \delta_L\right]
   \!+\! \lambda_R \mbox{Tr}\left[\Delta_R \delta_R\right]
   \!+\! \lambda_3 \mbox{Tr}\left[\tau_2 \Phi^T_1 \tau_2 \Phi_2\right]
  \\ &\
   + \lambda_4 \mbox{Tr}\left[\tau_2 \Phi^T_1 \tau_2 \Phi_1\right]
   + \lambda_5 \mbox{Tr}\left[\tau_2 \Phi^T_2 \tau_2 \Phi_2\right]
   + 3\lambda_S S^2 \Big|^{2}
\esp\ee
would lift the energy of the corresponding ground state. However, if $v_{R}$ and $\langle S\rangle =
v_S/\sqrt{2}$ are of the same order of magnitude and $\lambda_R$ and $\lambda_S$
are of opposite signs, this contribution can be kept small. In this
small part of the parameter space, one may have
values of $v_R$ clearly greater than $15$~TeV, so that one gets a decoupling
limit in which many particles become heavy due to the large
$v_R$ and $v_S$ VEVs, unless one invokes unforeseen fine-tuning effects.
In this limit, the $SU(2)_R$ gauge sector is easily beyond the reach of the LHC,
so that the light part of the entire LRSUSY spectrum may only include an
$SU(2)_R$ doubly-charged Higgs state, as
its mass is loop-suppressed relatively to the scale of left-right symmetry
breaking, in addition to the LSP, the dark matter candidate.

In the following, we assume
that the LSP belongs to the bidoublet higgsino state, so that extra neutralinos
and charginos are expected to be not too heavy as well. In this setup, a relic
density in agreement with the observations can be achieved if the higgsino
spectrum lies below 1~TeV. The only HL-LHC handle on the model
is then comprised of a signature made of a multi-leptonic system and missing
energy that emerges from resonant higgsino production (via
the $SU(2)_R$ gauge sector)~\cite{Chatterjee:2018gca}.
Other cosmologically-favored options could feature the lightest
right-handed sneutrino as a dark matter candidate. The spectrum does not
significantly differ here from the higgsino dark matter case, as to explain the
non-observation of any signal in the direct detection experiments and to avoid
dark matter over-abundance, one needs to rely on the existence of
co-annihilation channels~\cite{Chatterjee:2018gca}. This leads to a spectrum
including a rather light sneutrino in addition to a set of light higgsino
states. If the $W_R$ gauge boson is too heavy to be produced at the LHC, the
sneutrino signal will emerge either through multi-leptonic higgsino
cascade decays, as in the former higgsino dark matter case, or with a different
kinematic topology making it hard to access via traditional
searches~\cite{Ruiz:2017nip}. If the $W_R$ boson lies
instead within the reach of the HL-LHC, we should expect multi-leptonic signals
to originate from its decays into sleptons~\cite{Frank:2017tsm}.

The light doubly-charged Higgs state could then be the best probe of the model,
as suggested by recent studies on the sensitivity of future colliders in the
framework of Type-II seesaw and left-right models~\cite{deMelo:2019asm,Padhan:2019jlc,
Fuks:2019clu,Dev:2016dja,Dev:2018upe}.
However, in LRSUSY and in contrast with Type-II seesaw scenarios,
the couplings of the Higgs triplet are not determined by the
neutrino masses and mixings, as
the neutrino mass generation mechanism is comprised of a combination of Type-I
and Type-II seesaws. In this case, right-handed neutrinos indeed
get their mass through their interactions with the $\Delta_{R}$ weak triplet,
which then opens neutrino mass generation through the usual Type-I seesaw.

The light neutrino mass matrix is approximately given by
\be
(m_{\nu})_{il}=\frac{v_L^{2}}{2v_{R}}(\mathbf{Y}_{L}^{(2)})_{ij}(\mathbf{h}_{RR}^{-1})_{jk}(\mathbf{Y}_{L}^{(2)})^{\dagger}_{kl}\ ,
\ee
where $v_R$ ($v_u$) stands for the $SU(2)_R$ triplet (bidoublet) vacuum
expectation value. Even if we choose $\mathbf{h}_{RR}$ freely, there are still
enough freedom in the $\mathbf{Y}^{(2)}_{L}$ matrix so that it is possible to
reproduce the observed neutrino masses and mixings.

This means that the $SU(2)_R$ doubly-charged Higgs boson
could dominantly decay into a pair of same-sign tau leptons, taming the
sensitivity of the usually considered same-sign electron or muon channels.
In addition, as the doubly-charged Higgs state
belongs to an $SU(2)_R$ triplet and not an $SU(2)_L$ triplet as in the Type-II
case, the production mechanisms are related to different gauge
interactions. Moreover, the much heavier $SU(2)_R$ singly-charged Higgs bosons
tame the potential relevance of associated production down, again in contrast
with what typically occurs for $SU(2)_L$ triplets.

In the rest of this work, we focus on such LRSUSY scenarios in
which the Yukawa couplings responsible for the right-handed neutrino masses obey the same
generational hierarchy as for the other SM fermions, \ie, the coupling to the
third generation is the largest. The $SU(2)_R$ doubly-charged Higgs boson then
features a main decay mode into tau leptons, and it could be light without
violating any LHC constraint. Moreover, the lightest superpartners are the
neutral and charged bidoublet higgsinos, the lightest one being neutral and
a viable dark matter candidate. Such an LRSUSY configuration would be
experimentally the most challenging to observe, and therefore deserves the
present dedicated study.

\section{Collider phenomenology}
\label{sec:collider}
\subsection{Generalities}\label{subsec:gen}
As detailed in the previous section, the class of LRSUSY scenarios that we
consider focuses on setups in which the $SU(2)_R$ doubly-charged Higgs boson
($\dch$) is relatively light~\cite{Frank:2017tsm,
Chatterjee:2018gca}. In the minimal version of the model, its mass cannot be
much larger than 1~TeV~\cite{Basso:2015pka,Babu:2014vba}. The $\dch$ boson can
thus in principle be reachable at the 14 TeV LHC, even without the need of a
very high luminosity.

The derivation of the current limits on the $SU(2)_R$ doubly-charged
Higgs bosons requires some careful interpretation within the LRSUSY context. The
ATLAS collaboration has searched for doubly-charged scalars that decay to
electrons or muons~\cite{Aaboud:2017qph} after being produced through the
Drell-Yan mechanism (that is the dominant production mode in LRSUSY models).
The limits read
\be\mdch\gtrsim 650~{\rm GeV}\  , \ee
the exact bounds depending on the precise values of the $\dch$ branching ratios
in electrons and muons. Branching ratios as low as $BR(\dch\rightarrow
\ell^{\pm}\ell^{\pm})=4\%$ (for $\ell = e$ or $\mu$) are required to avoid most
bounds, so that $\mdch \gtrsim 350$~GeV.

The limits consequently become much weaker as soon as the $\dch$ state
dominantly decays into a pair of tau leptons. The corresponding search has been
performed by the CMS collaboration~\cite{CMS:2017pet}, both for a setup in which
the doubly-charged Higgs boson is pair produced and when it is produced in
association with a singly-charged Higgs boson. In LRSUSY models, singly-charged
Higgs bosons are typically a lot heavier than the doubly-charged ones, so that
they can be ignored from the discussion. In the case of a branching ratio
BR$(\dch\to\tau^{\pm}\tau^{\pm})=100\%$, the results lead to $\mdch\geq 396$~GeV
for $SU(2)_L$ doubly-charged Higgs bosons and slightly weaker bounds for
$SU(2)_R$ ones.

The most difficult combination would hence result from a scenario in which
BR$(H^{\pm\pm}_{R}\rightarrow \tau^{\pm}\tau^{\pm})=92\%$ and 
BR$(H^{\pm\pm}_{R}\rightarrow \mu^{\pm}\mu^{\pm})=$BR$(H^{\pm\pm}_{R}\rightarrow e^{\pm}e^{\pm})=4\%$.
In this case, we estimate that
\be \mdch \gtrsim 350\;\mathrm{GeV}.\ee

Under these circumstances, even HL-LHC operations are unlikely to be effective
in probing heavier $\dch$ bosons, especially if their mass gets close to 1~TeV.
This is the type of complicated scenarios that we are interested in in this
work. We hence aim at estimating the prospects of a potential high-energy
upgrade of the LHC at a center-of-mass energy of 27~TeV to probe scenarios in
which $\dch\to \tau^\pm\tau^{\pm}$ is the dominant decay mode.

We consider the all-hadronic
channel for which no dedicated study exists. In our analysis, we first design a
signal region {\bf SR1} targeting a signature arising from the production of a
pair of $\dch$ bosons decaying each into a di-tau system. In other words, we focus on
a final state featuring four hadronic tau leptons $\tau_h$. We however only
select
events in which three hadronic taus have been reconstructed, which guarantees
both sufficient signal rates after accounting for imperfect tau reconstruction, 
and a not too overwhelming SM background. Moreover, we additionally require some
missing transverse energy as it would stem from the tau decays.

Furthermore, we also explore how the subleading $\dch$ branching ratio
into same-sign electron or muon pairs could be used to get an extra handle on
the signal, as the corresponding signature is cleaner to reconstruct. We
define a second signal
region {\bf SR2} in which one selects events featuring a same-sign lepton (\ie,
electron or muon) pair and a di-tau system.

We therefore consider the following two signal regions,
\begin{equation*}\bsp
 &{\rm {\bf SR1\!:}~At~least~} 3 \tau_h, ~{\rm some~} \met; \\
 &{\rm {\bf SR2\!:}~At~least~} 2 \tau_h,
   ~{\rm 1~same\text{-}sign~lepton~pair,~some~} \met.
\esp\end{equation*}

In order to generate signal events at the HE-LHC for given benchmark scenarios,
we make use of the LRSUSY model implementation in the {\sc Sarah}
package~\cite{Staub:2013tta,Basso:2015pka}. This allows for both the computation
of the particle spectrum and branching ratios through {\sc SPheno~3}~\cite{
Porod:2011nf}, and for the translation of the model into the UFO format~\cite{
Degrande:2011ua} so that hard-scattering event generation could be achieved with
{\sc MG5\_aMC@NLO}~\cite{Alwall:2014hca}. For both signal and background, we
convolute leading-order matrix elements with the NNPDF 2.3 set of parton
distribution functions~\cite{Ball:2014uwa} and match the resulting events with
the parton showering and hadronization infrastructure of {\sc Pythia 8}~\cite{
Sjostrand:2014zea}. Subsequently, we implement the simulation of the detector
response with {\sc Delphes 3}~\cite{deFavereau:2013fsa}, that relies on the
anti-$k_T$ algorithm~\cite{Cacciari:2008gp} as implemented in the {\sc FastJet}
package~\cite{Cacciari:2011ma} for event reconstruction, and use the default
ATLAS detector parameterization. For a better
description of the background, we merge multipartonic matrix elements according
to the MLM procedure~\cite{Mangano:2006rw}, unless stated otherwise. Finally, in
order to obtain cosmologically-favored benchmark scenarios (see
Sec.~\ref{subsec:benchmarks}), we estimate the dark matter properties of our
scenarios with the {\sc MadDM} package~\cite{Ambrogi:2018jqj}.

\subsection{Benchmark scenarios}
\label{subsec:benchmarks}
In order to assess the sensitivity of the HE-LHC to the considered class of
LRSUSY scenarios, we select three representative benchmark configurations
{\bf BP1}, {\bf BP2} and {\bf BP3} featuring a different $SU(2)_R$
doubly-charged Higgs boson mass. The LSP is enforced to be part of the higgsino
bidoublets and its mass and properties are constrained to lead to a
cosmologically viable dark matter candidate. Its relic density is required
to agree with latest Planck data~\cite{Ade:2015xua}, which can be achieved
thanks to multiple co-annihilation processes among the six nearly
mass-degenerate higgsino-like neutralino and chargino states~\cite{
Chatterjee:2018gca}. The light part of the benchmark scenario spectra is presented in
Table~\ref{tab:bp_mass}, together with the relevant branching ratios of the
$\dch$ state.

We choose the Yukawa coupling matrix $h_{RR}$ to be diagonal and
yield branching
ratios BR$(H^{\pm\pm}_{R}\rightarrow \tau^{\pm}\tau^{\pm})=92\%$ and 
BR$(H^{\pm\pm}_{R}\rightarrow \mu^{\pm}\mu^{\pm})=$BR$(H^{\pm\pm}_{R}\rightarrow e^{\pm}e^{\pm})=4\%$.
As said above, this combination is the most challenging one with respect to the
searches. For instance, when the decay rate to one of the lighter leptons is
larger, stronger bounds immediately arise from the light lepton channels.

\begin{table}
  \renewcommand{\arraystretch}{1.6}
  \begin{center}
  \begin{tabular}{c c c c}
  & {\bf BP1} & {\bf BP2} & {\bf BP3} \\
  \hline
    $m_{W_R}$                &6550.5   &7486.2    &7486.2\\
    $m_{Z^{\prime}}$         &10993.2  &12563.6   &12563.6\\
    $\mdch$                  &875.0    &1016.4    &780.9\\
    $m_{\tilde\chi^0_1}$     &878.4    &803.7     &770.1\\
    $m_{\tilde\chi^0_2}$     &889.7    &812.1     &777.7\\
    $m_{\tilde\chi^0_3}$     &893.0    &815.3     &780.7\\
    $m_{\tilde\chi^0_4}$     &895.6    &817.6     &782.9\\
    $m_{\tilde\chi^0_5}$     &1032.2   &1043.5    &1048.0\\
    $m_{\tilde\chi^{\pm}_1}$ &886.4    &809.3     &775.1\\
    $m_{\tilde\chi^{\pm}_2}$ &893.5    &815.7     &781.2\\
    $\mdchg$                 &7413.2   &8412.9    &5619.2\\
  \hline
    BR($\dch\to\tau^\pm\tau^\pm$) &0.92   &0.92  &0.92\\
    BR($\dch\to\ell^\pm\ell^\pm$) &0.08   &0.08  &0.08\\
  \end{tabular}
  \caption{Relevant masses defining our three benchmark scenarios, given in GeV,
    and $\dch$ branching ratios. The table includes the masses of the $W_R$ and
    $Z'$ extra gauge bosons, the one of the $\dch$ state, those of the lightest
    neutralino $\tilde\chi^0_i$ (with $i = 1, 2, 3 ,4$), singly-charged charginos
    $\tilde\chi^\pm_i$ (with $i=1, 2$)
    and doubly-charged chargino $\tilde\chi^{\pm\pm}_1$.}
\label{tab:bp_mass}
\end{center}
\end{table}

\subsection{SR1: Investigating the triple-tau signature}

Our {\bf SR1} signal region focuses on events featuring at least three
reconstructed hadronic tau leptons and missing energy. SM backgrounds can arise
from QCD multi-jet processes and single boson production in association with
jets ($V$+jets, with $V\equiv W^{\pm}, Z$) when light jets are mis-tagged as
hadronic taus. We study, in our analysis, the dependence of the results on the
tau mis-tagging rate, that is allowed to vary between 1\% and 2\%. This choice
stems from the absence of any realistic mis-tagging rate expectation for a
potential future proton-proton collider at a center-of-mass energy of
27~TeV~\cite{Abada:2019ono}, and has been inspired by LHC capabilities for a
tau-tagging efficiency of 70\%~\cite{ATL-PHYS-PUB-2019-033,Sirunyan:2018pgf}.
Moreover, one expects subleading background contributions originating from $Zh$,
$hh$, $VV$ and
$VVV$ production in association with jets. For each component of the background
and all signal samples, the tau-tagging performances are included in the
evaluation of the rates presented in the following.

In order to avoid excessive multi-jet and $V$+jets background event generation
(to get numerical Monte Carlo uncertainties under control after accounting
for the small mis-tagging rates), we make use of the properties of the signal in
which quite hard tau jets with a large transverse momentum $p_T$ should
originate from the $\dch$ decays. We implement a set of generator-level cuts
(consistent with the subsequent analysis) while simulating those background
components, and hence require the transverse momentum of all (parton-level) jets
to satisfy $p_T > 40$~GeV.

\begin{table*}
  \setlength\tabcolsep{8pt}
  \renewcommand{\arraystretch}{1.2}
  \begin{center}
  \begin{tabular}{c||c|cccc}
    Process & Generator & Preselection & $H_T>1200$~GeV & $\met > 150$~GeV & $\mathcal{S}>0.3$\\
    \hline\hline
      {\bf BP1}    &0.251  &0.020 &0.013 &0.010 &0.006\\
      {\bf BP2}    &0.125  &0.011 &0.008 &0.007 &0.004\\
      {\bf BP3}    &0.430  &0.031 &0.017 &0.014 &0.008\\
    \hline
$pp\to VV$ + jets   &$3.08\times 10^5$ &$7.0\times 10^{-4}\ (0.0056)$ &$2.5\times 10^{-4}\ (0.002)$ &$\sim 10^{-5}$ &$\sim 10^{-5}$                   \\
$pp\to W^{\pm} +$ jets    &$1.31\times 10^7$  &0.065 (0.520) &0.014 (0.112) &0.003 (0.024) &$14.7\times 10^{-4}\ (0.011)$\\
$pp\to Z + $ jets   &$4.36\times 10^6$ &0.0206 (0.165)  &0.004 (0.032)  &$14.2\times 10^{-4}\ (0.011)$ &$5.0\times 10^{-4}\ (0.004)$                   \\
$pp\to$ jets        &$6.42\times 10^6$ &1.313 (10.504) &0.913 (7.304) &0.054 (0.432) &0.016 (0.128)
  \end{tabular}
  \caption{Cross sections, in fb, for the three benchmark signals and the
    different components of the SM background at various stages of the {\bf SR1}
    analysis. We present generator-level total rates (second column), as well as
    the reminding cross sections after the preselection (third column) and the
    various analysis cuts of Eq.~\eqref{eq:htcut}, Eq.~\eqref{eq:metcut} and
    Eq.~\eqref{eq:Scut} (last three columns). We consider a 70\% tau-tagging
    efficiency for a mis-tagging rate of 1\%. Results for a mis-tagging rate of
    2\% are given between parentheses (where relevant).}
  \label{tab:csec_sr1}
  \end{center}
  \renewcommand{\arraystretch}{1.0}
\end{table*}

In the multi-jet case,
we increase this selection criterion to $p_T>150$~GeV for the three hardest jets
and impose that the hadronic activity defined as the sum of the transverse
momenta of all reconstructed jets is greater than 1200~GeV. In practice,
we separately generate hard-scattering events for the $pp\to jjj $ and $pp\to
jjjj$ subprocesses and match them with parton showers, the di-jet case being
ignored given the need for the events selected in our analysis to feature at
least three hard jets faking tau leptons. Whilst in principle the overlap
between the tri-jet and tetra-jet samples should be removed by an appropriate
merging procedure, we instead directly combine those two samples for simplicity.
This is expected to yield a (conservative) over-estimation of the multi-jet
background.

For $V$+jets event generation, we similarly separately consider the $V+jj$ and
$V+jjj$ subprocesses and conservatively directly combine them. The overlap
between the two samples is however expected to be small, as any extra radiation
originating from a $V+jj$ final state is generally soft. All other background
components are treated as described in Sec.~\ref{subsec:gen}.

In the considered {\bf SR1} signal region, one preselects events by requiring
that they feature at least three hadronic taus $\tau_i$ (or jets faking taus)
with a transverse momentum
\be p_T(\tau_i) > 150~{\rm GeV}.\ee
We moreover veto the presence of any charged lepton with a
$p_T$ greater than 20~GeV and of any $b$-tagged jet with a $p_T$ greater than
25~GeV. Moreover, any system comprised of any two hadronic tau $(\tau_i,\tau_j)$
must have an invariant mass satisfying
\be
  m_{\tau_i\tau_j} > 200~{\rm GeV}~{\rm for~}i,j=1,2,3.
\ee
After this preselection, we impose that the scalar sum of the transverse
momentum of the three leading taus (including the potential jets faking taus)
satisfies
\be
  H_T = \sum_{i=1}^3 p_T(\tau_i) > 1200~{\rm GeV},
\label{eq:htcut}\ee
and require that the missing transverse energy fulfills
\be \label{eq:metcut}\met > 150~{\rm GeV}.\ee
Finally, in order to ensure a better rejection of the multi-jet background, we
require the event sphericity $\mathcal{S}$~\cite{Chen:2011ah}, computed from the three
selected taus, to be larger than 0.3,
\be  \mathcal{S} > 0.3\ . \label{eq:Scut}\ee

\begin{table}
  \setlength\tabcolsep{8pt}
  \renewcommand{\arraystretch}{1.2}
  \begin{center}
  \begin{tabular}{c|c|c}
    Scenario & Cut-and-count& Multivariate \\
    \hline
    {\bf BP1} &6.0   &1.95\\
    {\bf BP2} &12.38 &3.91\\
    {\bf BP3} &3.66  &1.14
  \end{tabular}
  \caption{Required luminosities, in ab$^{-1}$,  to obtain a $3\sigma$
    statistical significance with our cut-and-count (second column) and
    multi-variate (third column) analysis, for the three considered benchmark
    scenarios.} \label{tab:sign}
  \end{center}
  \renewcommand{\arraystretch}{1.0}
\end{table}

The corresponding cutflow is provided, for the three benchmark points and all
the components of the background, in Table~\ref{tab:csec_sr1}. As the tau-jets
arising from the on-shell decays of weak gauge and Higgs bosons are typically
softer than in the case of our signal, the corresponding multi-boson backgrounds
are drastically rejected already by our preselection cuts. The dominant
background components are therefore driven by jets faking hadronic tau leptons.
These can however be significantly reduced by the three extra cuts of
Eq.~\eqref{eq:htcut}, Eq.~\eqref{eq:metcut} and Eq.~\eqref{eq:Scut}, for
a moderate signal efficiency of about 50\%. As shown in Table~\ref{tab:sign},
about 6~ab$^{-1}$, 12.4~ab$^{-1}$ and 3.7~ab$^{-1}$ of data would be needed for
a $3\sigma$ statistical significance in the {\bf BP1}, {\bf BP2} and {\bf BP3}
cases, respectively, three luminosities that are well within the reach of the
HE-LHC, which is indeed expected to collect a luminosity as high as
15~ab$^{-1}$.

\begin{figure}
  \includegraphics[width=.32\textwidth]{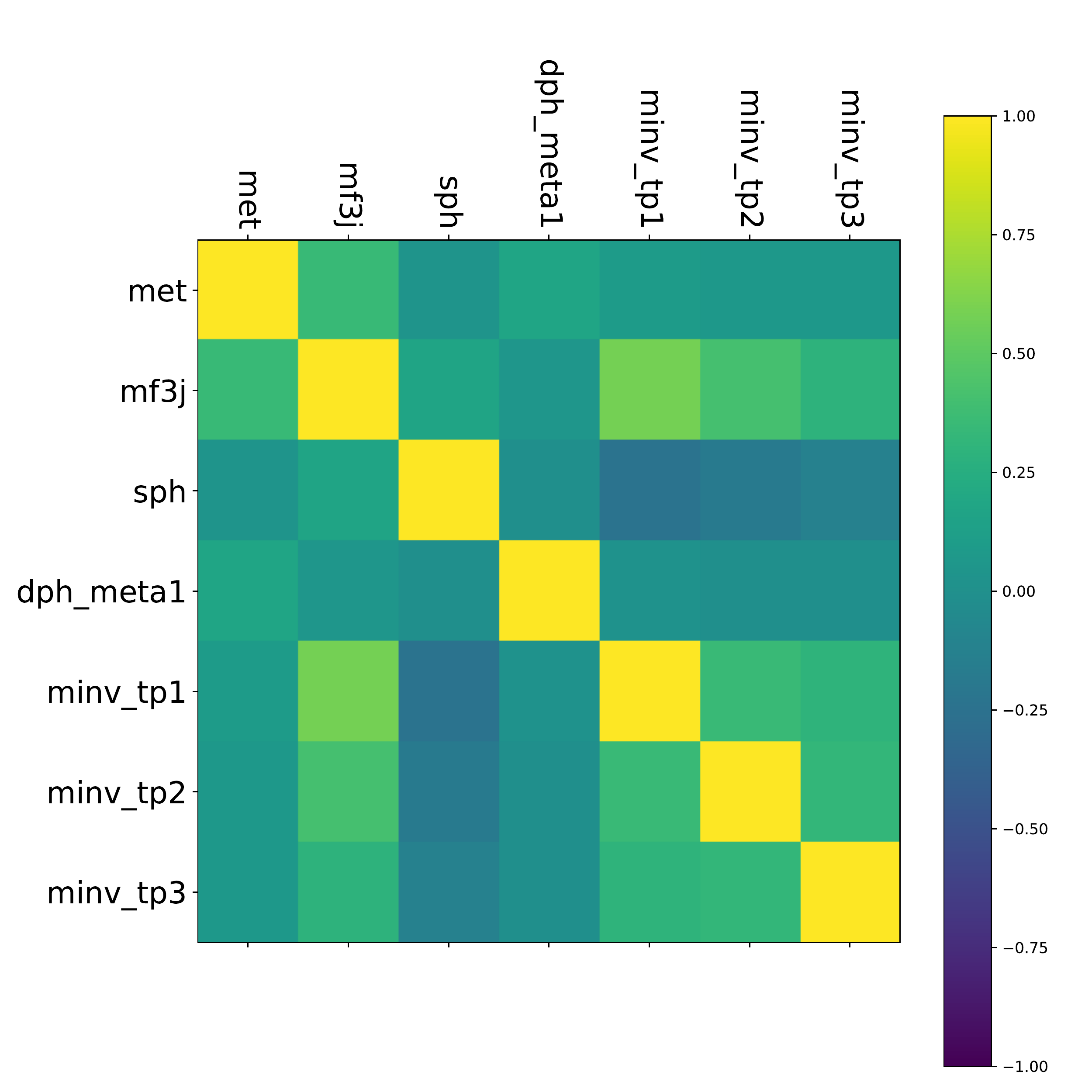}
  \caption{Correlations among the seven essential kinematic variables used in
    our multi-variate analysis of the {\bf BP1} doubly-charged Higgs boson
    signal. The labels of the $x$-axis and
    $y$-axis refer, from the first to the last variable, to 
    $\met$, $M_{\rm eff}$, $\mathcal{S}$, $\Delta\phi(\ptmiss,
    {\bf p}_T(\tau_1))$, $m_{\tau_1\tau_2}$, $m_{\tau_1\tau_3}$
    and $m_{\tau_2\tau_3}$, respectively.}
\label{fig:corr}
\end{figure}

\begin{figure}
  \includegraphics[width=.88\columnwidth]{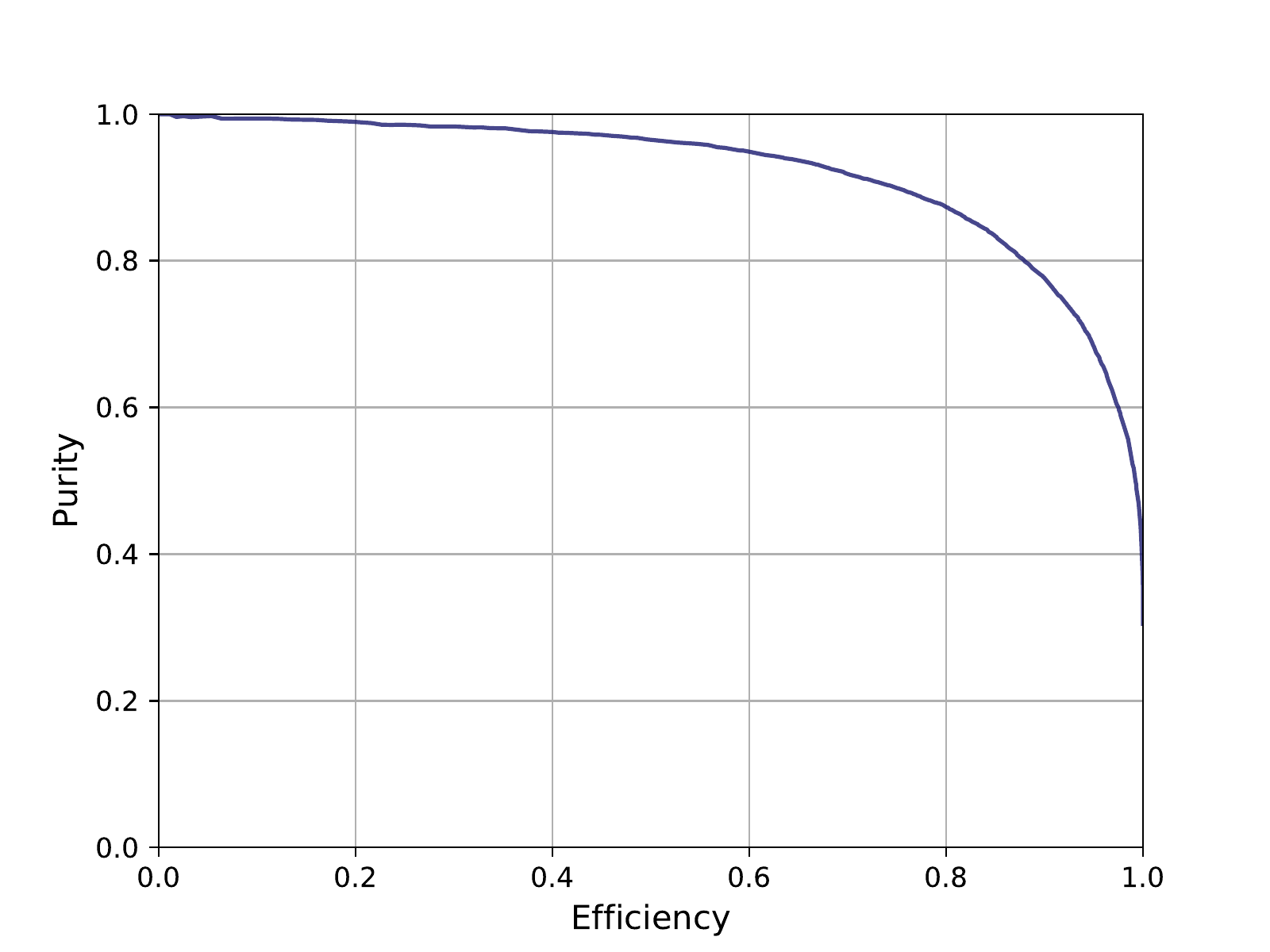}
  \includegraphics[width=.88\columnwidth]{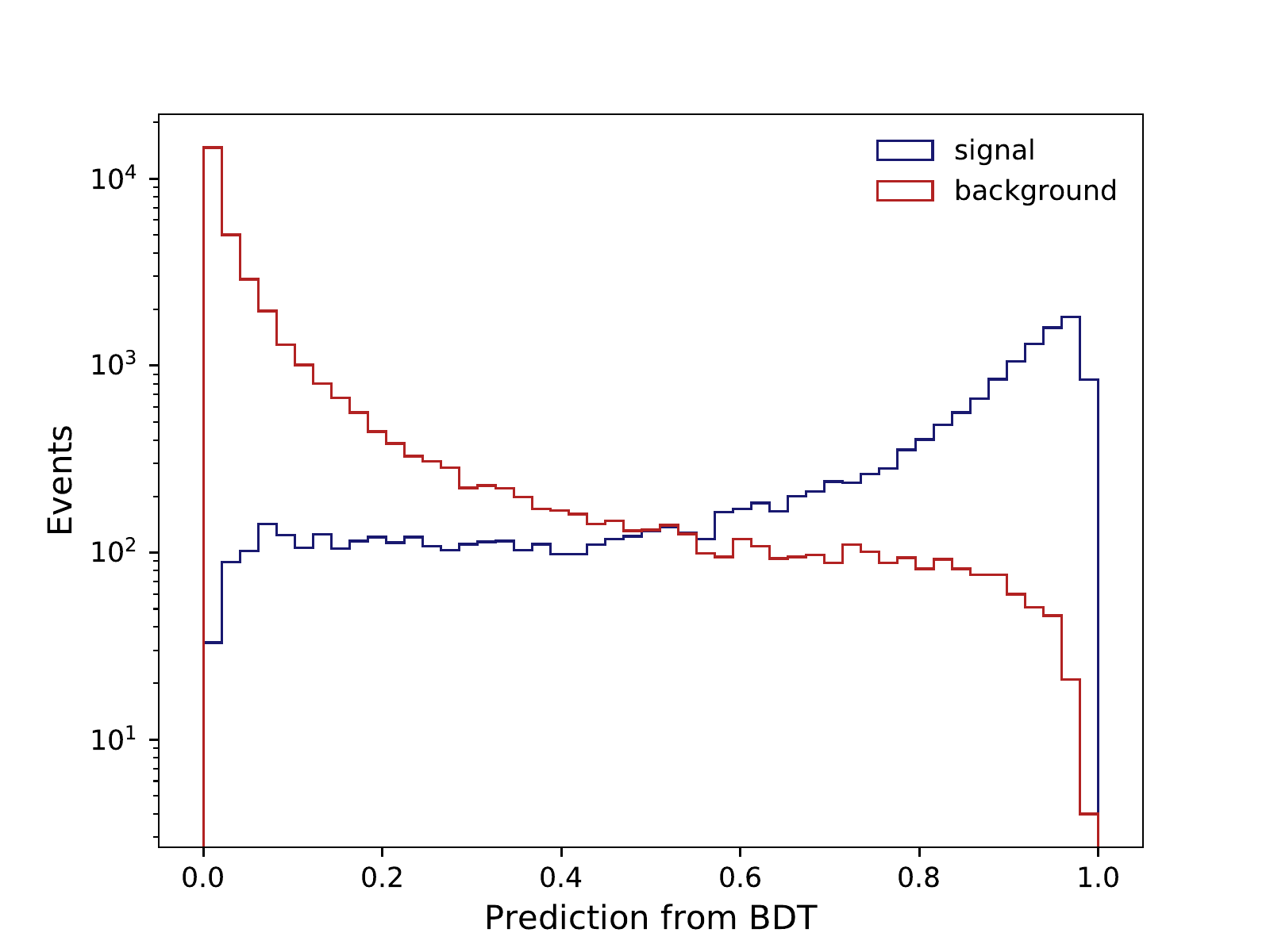}
  \caption{Properties of our BDT classifier in the context of the {\bf BP1}
    scenario. We present the obtained receiver operating characteristic (ROC)
    curve (upper panel), in which the efficiency measures the fraction of
    identified signal events that would pass a selection on the classifier and
    the purity denotes the ratio of identified signal events passing
    this selection to the total number of identified signal and background
    events. In the lower panel, we show the normalised distribution in the
    classifier for the signal (blue) and the background (red).}
\label{fig:mult_roc}
\end{figure}

Given the low sensitivity of the previously described cut-based analysis, we
 implement a complementarily multi-variate analysis based on a boosted decision
tree (BDT). We first preselect events as described in the cut-and-count analysis
and impose the $H_T$ selection of Eq.~\eqref{eq:htcut}. We then rely on seven
variables as inputs for our BDT classifier, namely the invariant mass of any
system made of any
pair of two of the three leading taus $m_{\tau_i\tau_j}$ (with $i,j=1, 2 ,3$),
the missing transverse energy $\met$, the effective mass $M_{\rm eff}$ defined
as
\be
 M_{\rm eff} = \sum_{i=1,3} p_T(\tau_i) + \met\ ,
\ee
the angular separation in azimuth between the missing transverse momentum
$\ptmiss$ and the leading tau $\Delta\phi(\ptmiss, {\bf p}_T(\tau_1))$
and the sphericity $\mathcal{S}$. We extract the HE-LHC sensitivity by
using the {\sc XGBoost} toolkit~\cite{Chen:2016btl}, employing the gradient
boosting method with a number of 1000 trees, a maximum depth of 4 and a learning
rate of 0.01. Our training set includes $80\%$ of the generated events, the
remaining events being then used for testing purposes. Moreover, we have
verified that our results were not affected by the addition of an extra variable
to the list of BDT inputs, both in the context of
basic quantities like the transverse momentum of any of the three leading taus
or their azimuthal separation, and in the context of more complex observables
like the aplanarity $A$~\cite{Chen:2011ah}, the missing energy significance
$\met/\sqrt{H_T}$, the $\met/M_{\rm eff}$ ratio, or the relative $p_T$ of the
third tau with respect to the two leading ones $y_{23}$,
\be y_{23} = \frac{p^2_T(\tau_3)}{\big[p_T(\tau_1) + p_T(\tau_2)\big]^2} \ . \ee
Including any extra variable on top of the seven above-mentioned ones has indeed
only been found to increase the correlations. While the relative relevance of
the variables varies from benchmark to benchmark, the level of correlations is
maintained in each case to a low level, as illustrated in
Fig.~\ref{fig:corr} for the {\bf BP1} scenario.

As evident from Fig.~\ref{fig:mult_roc} for the {\bf BP1} scenario, our BDT
classifier is quite efficient in identifying signal events whilst rejecting
background events, as signal efficiencies larger than 80\% can be obtained
together with high background rejection rates. In the upper panel of the
figure, we present
the receiver operating characteristic (ROC) curve of the algorithm. The
area under the curve (AUC) is a good indicator of the algorithm
effectiveness as it should approach 1 for well-performing methods. It is found
to be of about 0.92 for the {\bf BP1} scenario, whilst similar results are
obtained for the {\bf BP2} and {\bf BP3} scenarios, with AUC of 0.88 and 0.82
respectively. In the lower panel of Fig.~\ref{fig:mult_roc},
we present the signal and background distributions (in the case of the {\bf BP1}
scenario) according to the BDT classifier, which reinforces the illustration of
its good discriminating power. This largely impacts the sensitivity of the
HE-LHC to the model, the resulting significance factors being importantly improved relatively to
the cut-based analysis. This is complementarily demonstrated in
Table~\ref{tab:sign} in which
the HE-LHC luminosity needed to reach a $3\sigma$ statistical significance is
presented for each of our three representative scenarios. In all three cases,
this luminosity is found three times smaller than for the cut-based analysis.

\subsection{SR2: Investigating the di-tau plus di-lepton signature}

\begin{table}
  \setlength\tabcolsep{8pt}
  \renewcommand{\arraystretch}{1.2}
  \begin{center}
  \begin{tabular}{c||cc}
 Process & C1 & C2 \\
	  \hline\hline 
	  {\bf BP1} &0.094  &0.003 \\ 
	  {\bf BP2} &0.050 &0.002 \\ 
	  {\bf BP3} &0.150 &0.005 \\
	  \hline
	  $pp\to VV$ + jets &0.238 (0.952) &$10^{-4}$ ($4\times 10^{-4}$) \\
	  
  \end{tabular}
  \caption{Cross sections in fb, for the signal as well as the most dominant background process after imposing the selection criteria for the
	  {\bf SR2} signal region. We follow the same convention as in Table~\ref{tab:csec_sr1}.} \label{tab:csec_sr2}
  \end{center}
  \renewcommand{\arraystretch}{1.0}
\end{table}

\begin{figure}
  \includegraphics[height=8cm,width=9cm]{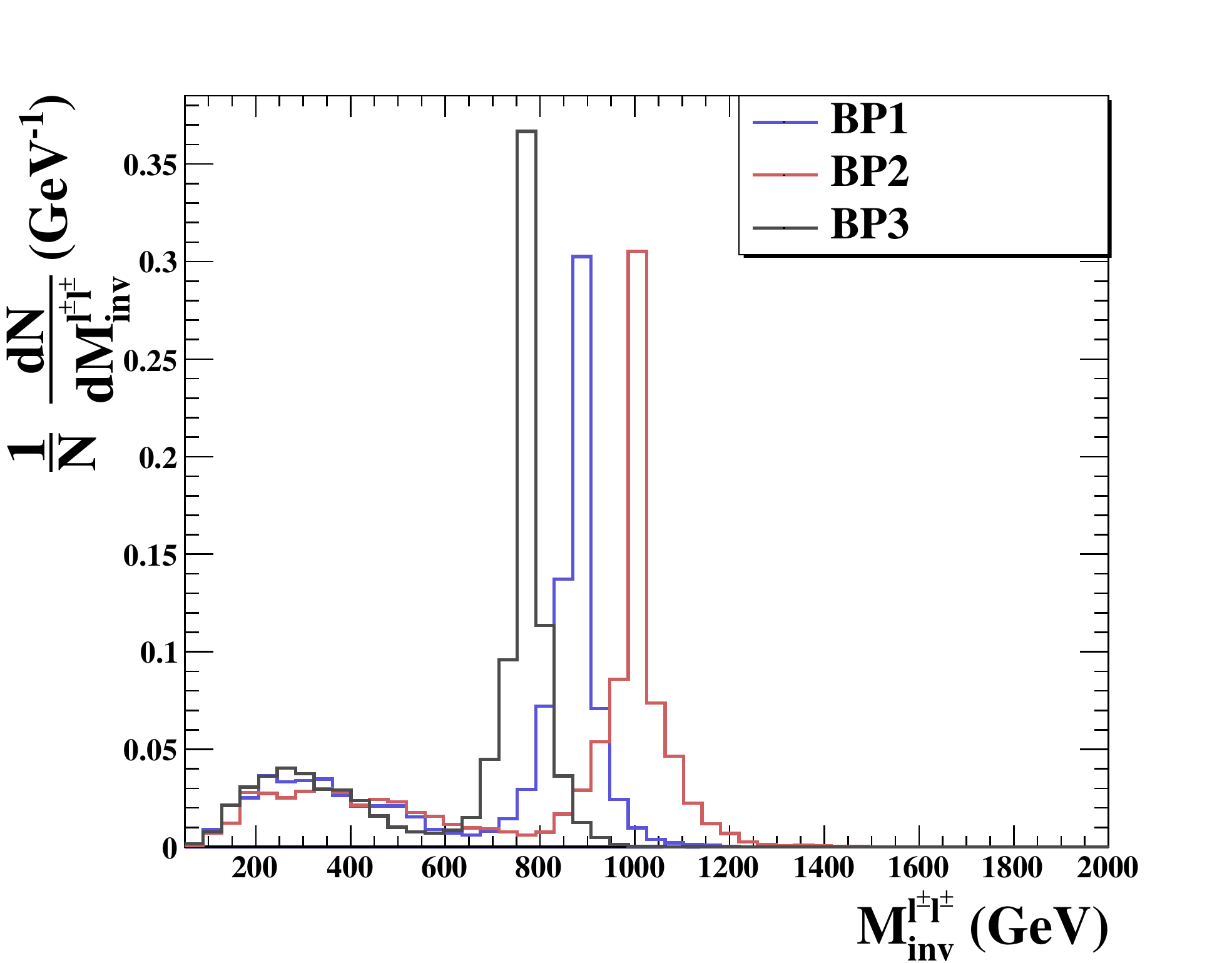}
  \caption{Distribution in the invariant mass of the same-sign-same-flavor
    lepton pair originating from the $\dch$ decay for the three chosen benchmark
    points, after imposing the {\bf SR2} selection.} \label{fig:minv_lp}
\end{figure}

In our three benchmark scenarios, the $\dch$ branching ratio into a pair of
same-sign electrons or muons has been fixed to 8\% (see
Table~\ref{tab:bp_mass}), which is small enough to evade the current LHC
constraints originating from doubly-charged Higgs boson searches~\cite{
Frank:2017tsm}. As a result, any handle on the model relying on the
doubly-charged Higgs boson decay into a same-sign-same-flavor lepton pair is
associated with a quite small signal cross section. However, the requirement for
the presence of these two leptons along two hard hadronic taus ensures that there is not much
background surviving the selection. 

In a first C1 preselection, we impose that events contain at least two
hadronic taus with a transverse momentum $p_T > 150$~GeV and we additionally
constrain the invariant mass of the system made of the two hardest taus to be
larger than 200~GeV. We moreover veto the presence of $b$-jets. We then add a C2
selection in which we focus on events featuring a pair of same-sign electrons or
muons with $p_T > 50$~GeV. With these criteria, the SM background contribution is rendered
negligible, with a cross section of about $10^{-4}$ fb after including
the $Zh$, $VV$ + jets, $VVV$ and $V$ + jets components of the background. After
the lepton requirements C2, only $VV$ + jets events in fact contribute to the background.
The signal rates in the 
context of the three considered benchmark scenarios and the most dominant background channel 
are presented in Table~\ref{tab:csec_sr2}. We equivalently obtain an
almost background-free environment for a handful of signal events for the
expected 15~ab$^{-1}$ luminosity of the HE-LHC.

As a consequence, the {\bf SR2}
analysis may provide a complementary handle on the signal, relatively to the
{\bf SR1} one, with an enormous advantage in the fact that the doubly-charged
Higgs-boson mass can be reconstructed from the analysis of the properties of the
pair of same-sign leptons (that moreover consists in a smoking gun signal for a
doubly-charged Higgs boson). This mass reconstruction can be quite precise
despite the detector effects, as illustrated in Fig.~\ref{fig:minv_lp}. In this
figure, we present the invariant-mass spectrum of the di-lepton-pair for all
three benchmark scenarios. In each case, the distribution exhibits a clear peak
located right at the doubly-charged Higgs-boson mass.

Apart from the two signal regions defined in this work, one can also build an
analysis targeting a signature stemming from the production of multiple higgsino states.
Altogether, the considered class of scenarios features spectra in which four
neutralino and two chargino states are nearly mass degenerate and sitting at the
lighter part of the LRSUSY model particle spectrum. However, as a consequence,
any SM jets and/or leptons that may arise from higgsino production and decay is
expected to be too soft to be detected. The standard probe to such scenarios
consists thus of the monojet channel, that is at least promising in MSSM-like
scenarios. In this last case, higgsino masses of 500~GeV can be reached at the
HE-LHC~\cite{Han:2018wus}. Owing to a richer higgsino sector in the LRSUSY
framework, one can expect a larger signal cross section for a fixed mass, so
that heavier higgsino states could conversely be probed. However,
the lower limit on the higgsino mass, so that we could
obtain a viable dark matter candidate is of about 700~GeV~\cite{
Chatterjee:2018gca}. The corresponding monojet rates have been found to be
subsequently too low to lead to any observable signal with 15~ab$^{-1}$ of
HE-LHC luminosity.

\section{Summary and Conclusion}
\label{sec:conclusion}
We have analyzed the sensitivity of a high-energy upgrade of the LHC expected
to operate at a center-of-mass energy of 27~TeV (\ie, the HE-LHC) to a class of
left-right supersymmetric scenarios favored by dark matter, with a relic density
as measured by the Planck collaboration originating from the co-annihilations of
multiple higgsino states of about 700~GeV. In a minimal LRSUSY setup where the
stabilization of the vacuum occurs through radiative corrections, the
doubly-charged Higgs boson mass is loop-suppressed relatively to
the rest of the $SU(2)_R$ sector, so that we expect it to be the first
manifestation of the model at colliders.
As previous experimental limits on this state are obtained
by assuming a pair-production mode followed by a decay into a same-sign pair of
electrons or muons, we focus on the still phenomenologically viable option where
the doubly-charged Higgs boson decays almost exclusively into tau leptons. We
explore this possibility and estimate the chances to observe such an LRSUSY
scenario at the HE-LHC.

We consider in particular two signatures, namely a first one where the pair-produced
doubly-charged Higgs states decay into tau leptons, and a second one in which
one of them is assumed to decay into an electron or a muon pair. For the former
case, we focus on the production of at least three hard hadronic tau leptons.
The SM background is mostly comprised of multi-jet and vector-boson-plus-jets
events in which QCD jets are faking tau leptons. We have implemented a series of
cuts which not only lead to a good
significance at high luminosities, but that can also serve as a basis for a
multi-variate analysis relying on boosted decision trees. In this case, three
times less luminosity could be required to observe a 3$\sigma$ signal (which
would occur thus at an early HE-LHC stage). For our latter analysis, we position
ourselves in an almost background-free environment by investigating a di-tau
plus same-sign di-lepton signature. Whereas signal cross sections are expected
to be small, the large HE-LHC luminosity makes this analysis a nice
complementary handle on the previously considered LRSUSY signal. Moreover, the presence of
the two first or second generation leptons guarantees the reconstruction of the
doubly-charged Higgs boson mass. In contrast, any signal that could arise from
the large number of light higgsino states has been found not to give any hope
for a discovery, as the corresponding monojet cross sections are way too small.

In summary, in the minimal left-right supersymmetry setup in which one relies on
radiative corrections to stabilize the vacuum configuration,
we expect the first signal of new physics to arise from the
doubly-charged Higgs boson. However, it may hidden as decaying mainly into a
pair of same-sign tau leptons. We have shown that at the HE-LHC, we may see a
signal for doubly-charged Higgs boson masses ranging up to around 1~TeV, this
upper limit being theoretically motivated as it requires to push the left-right
symmetry breaking scale in a way
that is only possible in a very small part of the parameter space.
In addition to the doubly-charged Higgs boson, the model includes a dark matter
candidate that cannot be too heavy. The latter will however hardly provide any
clear signal without the help of the $SU(2)_R$ gauge sector, as shown in previous
work.

If the doubly-charged Higgs boson fails to materialize at the HE-LHC, we may
conclude that the LRSUSY vacuum has likely to be
stabilized by some other mechanism than by loop corrections.
Models employing an even larger Higgs sector or those breaking the
$R$-parity spontaneously may have a tree-level contribution to the
doubly-charged Higgs boson mass, which could thus be larger.

\section*{Acknowledgements}
MF thanks NSERC for support through grant number SAP105354. SKR  acknowledges financial support from the Department of Atomic Energy, Government of India, for the
Regional Centre for Accelerator-based Particle Physics (RECAPP), Harish-Chandra Research Institute.
HW acknowledges the support from Magnus Ehrnrooth Foundation and STFC Rutherford International Fellowship (funded through MSCA-COFUND-FP, grant number 665593).
\bibliography{lrsusyhelhc}
\end{document}